\documentclass{elsart5p}

\usepackage{graphics}

\journal{Physics Letters B}

\begin{document}
\renewcommand{\textfraction}{0.00000000001}
\renewcommand{\floatpagefraction}{1.0}
\begin{frontmatter}
\title{Coherent photoproduction of $\eta$-mesons off $^3$He - search for
$\eta$-mesic nuclei}

\author[Basel]{F.~Pheron},
\author[Mainz]{J.~Ahrens},             
\author[Glasgow]{J.R.M.~Annand},          
\author[Mainz]{H.J.~Arends},
\author[Kent]{K.~Bantawa},
\author[Mainz]{P.A.~Bartolome},            
\author[Bonn]{R.~Beck},         
\author[Petersburg]{V.~Bekrenev},
\author[Giessen]{H.~Bergh\"auser},            
\author[Basel]{B.~Boillat},             
\author[Pavia]{A.~Braghieri},           
\author[Edinburgh]{D.~Branford},            
\author[Washington]{W.J.~Briscoe},           
\author[UCLA]{J.~Brudvik},             
\author[Lebedev]{S.~Cherepnya},           
\author[Washington]{B.~Demissie},
\author[Basel]{M.~Dieterle},             
\author[Mainz,Glasgow,Washington]{E.J.~Downie},            
\author[Giessen]{P.~Drexler},              
\author[Edinburgh]{D.I.~Glazier},           
\author[Mainz]{E.~Heid}, 
\author[Lebedev]{L.V. Fil'kov}, 
\author[Sackville]{D.~Hornidge}, 
\author[Glasgow]{D.~Howdle},            
\author[Mainz]{O.~Jahn},
\author[Basel]{I.~Jaegle},
\author[Edinburgh]{T.C.~Jude},                
\author[Lebedev,Mainz]{V.L.~Kashevarov},
\author[Basel]{I. Keshelashvili},        
\author[Moskau]{R.~Kondratiev},          
\author[Zagreb]{M.~Korolija},  
\author[Basel,Giessen]{M.~Kotulla},          
\author[Petersburg]{A.~Kulbardis},
\author[Petersburg]{S.P. Kruglov}, 
\author[Basel]{B.~Krusche},\ead{Bernd.Krusche@unibas.ch}          
\author[Moskau]{V.~Lisin},               
\author[Glasgow]{K.~Livingston},          
\author[Glasgow]{I.J.D.~MacGregor},       
\author[Basel]{Y.~Maghrbi},
\author[Glasgow]{J.~Mancell},  
\author[Kent]{D.M.~Manley}, 
\author[Washington]{Z.~Marinides},           
\author[Mainz]{M.~Martinez},
\author[Glasgow]{J.C.~McGeorge}, 
\author[Glasgow]{E.~McNicoll},          
\author[Zagreb]{D.~Mekterovic},          
\author[Giessen]{V.~Metag},
\author[Zagreb]{S.~Micanovic},
\author[Sackville]{D.G.~Middleton},
\author[Pavia]{A.~Mushkarenkov},               
\author[UCLA]{B.M.K.~Nefkens},         
\author[Bonn]{A.~Nikolaev},   
\author[Giessen]{R.~Novotny},
\author[Basel]{M.~Oberle},
\author[Mainz]{M.~Ostrick},
\author[Mainz,Washington]{B.~Oussena},
\author[Pavia]{P.~Pedroni},             
\author[Moskau]{A.~Polonski},            
\author[UCLA]{S.N.~Prakhov}, 
\author[Glasgow]{J.~Robinson},                    
\author[Glasgow]{G.~Rosner},              
\author[Basel,Pavia]{T.~Rostomyan},           
\author[Mainz]{S.~Schumann},
\author[Edinburgh]{M.H.~Sikora},            
\author[Catholic]{D.I.~Sober},               
\author[UCLA]{A.~Starostin},           
\author[Zagreb]{I.~Supek},               
\author[Giessen]{M.~Thiel},        
\author[Mainz]{A.~Thomas},              
\author[Mainz]{M.~Unverzagt},                       
\author[Edinburgh]{D.P.~Watts},
\author[Basel]{D.~Werthm\"uller},
\author[Basel]{L.~Witthauer}, 
\author[Basel]{F.~Zehr}

\address[Basel] {Department of Physics, University of Basel, Ch-4056 Basel, Switzerland}
\address[Mainz] {Institut f\"ur Kernphysik, University of Mainz, D-55099 Mainz, Germany}
\address[Glasgow] {SUPA, School of Physics and Astronomy, University of Glasgow, G12 8QQ, United Kingdom}
\address[Kent] {Kent State University, Kent, Ohio 44242, USA}
\address[Bonn] {Helmholtz-Institut f\"ur Strahlen- und Kernphysik, University Bonn, D-53115 Bonn, Germany}  
\address[Petersburg] {Petersburg Nuclear Physics Institute, RU-188300 Gatchina, Russia}
\address[Giessen] {II. Physikalisches Institut, University of Giessen, D-35392 Giessen, Germany}
\address[Pavia] {INFN Sezione di Pavia, I-27100 Pavia, Pavia, Italy}
\address[Edinburgh] {School of Physics, University of Edinburgh, Edinburgh EH9 3JZ, United Kingdom}
\address[Washington] {Center for Nuclear Studies, The George Washington University, Washington, DC 20052, USA}
\address[UCLA] {University of California Los Angeles, Los Angeles, California 90095-1547, USA}
\address[Lebedev] {Lebedev Physical Institute, RU-119991 Moscow, Russia}
\address[Sackville] {Mount Allison University, Sackville, New Brunswick E4L1E6, Canada}
\address[Moskau] {Institute for Nuclear Research, RU-125047 Moscow, Russia}
\address[Zagreb] {Rudjer Boskovic Institute, HR-10000 Zagreb, Croatia}
\address[Catholic] {The Catholic University of America, Washington, DC 20064, USA}

\begin{abstract}
Coherent photoproduction of $\eta$-mesons off $^3$He, i.e. the reaction
$\gamma ^3\mbox{He}\rightarrow \eta ^3\mbox{He}$, has been investigated 
in the near-threshold region. The experiment was performed at the Glasgow tagged photon 
facility of the Mainz MAMI accelerator with the combined Crystal Ball - TAPS
detector. Angular distributions and the total cross section were measured
using the $\eta\rightarrow \gamma\gamma$ and 
$\eta\rightarrow 3\pi^0\rightarrow 6\gamma$ decay channels. 
The observed extremely sharp rise of the cross section at threshold and the 
behavior of the angular distributions are evidence for a strong $\eta {^3\mbox{He}}$ final 
state interaction, pointing to the existence of a resonant state. The search 
for further evidence of this state in the excitation function of $\pi^0$-proton 
back-to-back emission in the $\gamma ^3\mbox{He}\rightarrow \pi^0 pX$ reaction 
revealed a very complicated structure of the background and could not support
previous conclusions. 
 \end{abstract}
\end{frontmatter}

\newpage

\section{Introduction}

The interaction of mesons with nuclei is a major source
for our understanding of the strong interaction. For 
long-lived mesons, like charged pions or kaons, secondary beams can be used
for detailed studies of elastic and inelastic reactions, revealing the
relevant potentials. However, most mesons are short-lived so that their
interaction with nucleons can only be studied in indirect ways, making use of
final-state interactions (FSI). The general idea is to produce the
mesons with some initial reaction in a nucleus and then study their interaction
with the same nucleus. 

It is much discussed whether the strong interaction allows the formation of 
quasi-bound meson-nucleus states. So far, all known meson-nucleus bound 
states involve at least partly the electromagnetic interaction. Pionic atoms 
are well established. More recently deeply bound pionic states have also
been reported \cite{Geissel_02}, but in this case the binding results 
from the superposition of the repulsive s-wave $\pi^-$-nucleus interaction
with the attractive Coulomb force. Neutral mesons on the other hand 
could form quasi-bound states only via the strong interaction.  
The meson-nucleus interaction for slow pions is much too weak 
to produce quasi-bound states, but the situation may be different for $\eta$,
$\eta '$, and $\omega$-mesons. 

The case of $\eta$-mesons is special because the
threshold region of $\eta$-production is dominated by an $s$-wave
resonance \cite{Krusche_95,Krusche_97}, the S$_{11}$(1535), which couples 
very strongly to $N\eta$ (branching ratio $\approx$50\% \cite{PDG}).
As a consequence FSI in nuclei are important; $\eta$-mesons are absorbed in 
nuclei with typical cross sections around 30~mb, basically independent of 
their kinetic energy $T$ over a wide range of $T$ from a few MeV to 1 GeV 
\cite{Roebig_96,Mertens_08}.  
First hints for the possible existence of bound $\eta$-nucleus states came 
from the analysis of the $\eta N$ scattering length that characterizes the
potential at low kinetic energies. Already in 1985 Bhalerao 
and Liu \cite{Bhalerao_85} reported an attractive $s$-wave $\eta N$ interaction
from a coupled channel analysis of pion-induced reactions, yielding
scattering lengths $\alpha_{\eta N}$ with real parts between 0.27~fm and 
0.28~fm and imaginary parts between 0.19~fm and 0.22~fm. Shortly afterwards, 
Liu and Haider \cite{Liu_86} pointed out that this interaction might 
lead to the formation of quasi-bound $\eta$-nucleus states for $A >$~10. 
Experimental evidence for such `heavy' $\eta$-mesic nuclei has  
been searched for in pion-induced reactions on nuclei in the oxygen region 
\cite{Chrien_88,Johnson_93}, but those experiments did not produce conclusive 
evidence. More recently, Sokol and co-workers \cite{Sokol_99,Sokol_08}
claimed evidence for the formation of $\eta$-mesic nuclei from 
bremsstrahlung-induced reactions on $^{12}$C 
\begin{equation}
\gamma + ^{12}\mbox{C}\rightarrow 
p(n) + ^{11}_{\eta}\mbox{B}(^{11}_{\eta}\mbox{C})\rightarrow
\pi^+ + n + X\;\;,
\end{equation}
where the $n\pi^+$ pairs were detected in the final state. In this type of
experiment, the $\eta$-meson is produced in quasi-free kinematics on a nucleon 
($p,n$), which takes away the largest part of the momentum and the
$\eta$ is almost at rest in the residual $A=11$ nucleus. If a
quasi-bound state is produced, the $\eta$-meson has a large chance to be 
re-captured by a nucleon into the S$_{11}$ excitation, which may then decay 
into a pion-nucleon back-to-back pair. Such pairs were searched for in the
experiment, however, background from quasi-free pion production must be
considered. Sokol and Pavlyuchenko \cite{Sokol_08} claim an enhancement above
this background for certain kinematic conditions.

The topic gained much new interest after precise low-energy data for the 
photoproduction of $\eta$-mesons off the proton \cite{Krusche_95}, deuteron 
\cite{Krusche_95a,Hoffmann_97,Weiss_01,Weiss_03} 
and helium nuclei 
\cite{Hejny_99,Hejny_02} became available and refined model analyses of the 
scattering length were done by many groups (see \cite{Arndt_05} for a summary). 
The results for the imaginary part are rather stable, most cluster between 
0.2~fm and 0.3~fm. The real part, which determines the existence of quasi-bound 
states, is much less constrained. It runs all the way from a negative value 
of $-$0.15~fm to numbers close to and even above $+$1~fm. However, most of the 
more recent analyses prefer large values above 0.5~fm, which has raised 
controversial discussions about the possible existence of very light mesic nuclei 
and prompted theoretical studies of the $\eta$-interaction with $^2$H, $^3$H, $^3$He, 
and $^4$He systems 
\cite{Ueda_91,Ueda_92,Wilkin_93,Rakityanski_95, Rakityanski_96,Green_96,Scoccola_98,Shevchenko_00,Grishina_00,Garcilazo_01} 

Experimental evidence for light $\eta$-mesic nuclei has mostly been
searched for in the threshold behavior of $\eta$-production reactions.
The idea is that quasi-bound states in the vicinity of the production
threshold will give rise to an enhancement of the cross section relative to 
the expectation for phase-space dependence. 
Many different hadron induced reactions have been studied in view of such
threshold effects. Already the measurement of pion induced $\eta$-production off
$^3$He in the reaction $\pi^- + {^3\mbox{He}}\rightarrow\eta + t$ at LAMPF
\cite{Peng_87,Peng_89} revealed production cross sections significantly larger 
than DWIA predictions. Subsequently, different nucleon-nucleon and nucleon-deuteron
reactions, in particular
$pp\rightarrow pp\eta$ \cite{Calen_96,Smyrski_00,Moskal_04}, 
$np\rightarrow d\eta$ \cite{Plouin_90,Calen_98}, 
$pd\rightarrow\eta ^3\mbox{He}$ \cite{Mayer_96},
$dp\rightarrow\eta ^3\mbox{He}$ \cite{Smyrski_07,Mersmann_07,Rausmann_09}, 
$dd\rightarrow\eta ^4\mbox{He}$ \cite{Wronska_05},
$\vec{d}d\rightarrow \eta ^4\mbox{He}$ \cite{Willis_97,Budzanowski_09}, 
and $pd\rightarrow pd\eta$ \cite{Hibou_00}
have been studied. Interesting threshold effects have been found in most of them,
but in particular the $pd\rightarrow\eta  ^3\mbox{He}$ \cite{Mayer_96} and
$dp\rightarrow\eta  ^3\mbox{He}$ reactions \cite{Smyrski_07,Mersmann_07,Rausmann_09}
show an extremely steep rise at threshold, implying a very large $\eta^3\mbox{He}$
scattering length. Wilkin and collaborators \cite{Wilkin_07} have argued from an
analysis of the angular distributions that, not only the magnitude of the $s$-wave 
amplitude falls rapidly in the threshold region, but that its phase also varies
strongly, which supports the idea of a quasi-bound or virtual ~$_{\eta}^3\mbox{He}$ 
state very close to the threshold. 

If such states do exist, they should show up as threshold enhancements
independently of the initial state of the reaction. Photoproduction of
$\eta$-mesons from light nuclei is a very clean tool for the preparation of the
$\eta$-nucleus final state with small relative momenta but, due to the 
much smaller electromagnetic cross sections, sensitive threshold 
measurements have been sparse until now. A particular problem is that the $\eta$-mesons
have to be produced coherently off the target nuclei. As a consequence of the 
results from photoproduction of $\eta$-mesons off light nuclei, it is now well 
understood that the threshold production is dominated
by an isovector spin-flip amplitude exciting the S$_{11}$(1535) resonance 
(see \cite{Krusche_03} for a summary). Therefore, the coherent cross section is
very small for the isoscalar deuteron and practically forbidden for the isoscalar-scalar
$^4$He nucleus, where only higher partial waves could contribute. Among the light targets 
only the ($I=1/2$,$J=1/2$) $^3$H and $^3$He nuclei have reasonably large cross sections 
for the $\gamma A\rightarrow A\eta$ reaction.

The $^3$He system was previously investigated with photon-induced reactions 
by Pfeiffer et al. \cite{Pfeiffer_04}. Possible evidence for the formation of a 
quasi-bound state was reported from the behavior of two different reactions. 
The coherent $\eta$-photoproduction $\gamma ^3\mbox{He}\rightarrow \eta^3\mbox{He}$ 
showed a strong threshold enhancement with angular distributions much more isotropic
in the threshold region than expected from the nuclear form factor. These are
indications for strong FSI effects. Furthermore, in an approach similar 
to the Sokol experiment \cite{Sokol_08}, the excitation function for $\pi^0-p$
pairs emitted back-to-back in the $\gamma ^3\mbox{He}$ center-of-momentum (c.m.)
system was investigated.
The difference between the excitation functions for opening angles between 
170$^{\circ}$ - 180$^{\circ}$ (almost back-to-back) and 150$^{\circ}$ - 170$^{\circ}$
(background from quasi-free pion production) showed a narrow peak around the
$\eta$-production threshold. Such a peak is expected to arise when quasi-bound
$\eta$-mesons are re-captured into a nuclear S$_{11}$ excitation with subsequent
decay into the pion-nucleon channel. However, the statistical quality of the 
measurements was limited and it was pointed out by several authors 
\cite{Sibirtsev_04,Hanhart_05} that the data do not prove the existence of a 
quasi-bound state. 

The present experiment aimed at a measurement with significantly improved 
statistical quality and better control of systematic effects. This was 
achieved by using a detector system with much larger solid-angle coverage.
Apart from counting statistics this is important for two aspects. It allows
a measurement of not only the $\eta\rightarrow 2\gamma$ decays but also the 
$\eta\rightarrow 3\pi^0\rightarrow 6\gamma$ decay chain. Comparison of the two
results helps to estimate systematic effects from the detection efficiency.
Even more important, the coincident detection of recoil nucleons from
non-coherent production processes becomes much more efficient and can be used 
to suppress this most important background.

\section{Experiment and analysis}

The experiment was performed at the tagged photon beam of the Mainz MAMI accelerator 
\cite{Herminghaus_83,Kaiser_08}. The electron beam of 1508 MeV was used to produce
bremsstrahlung photons in a copper radiator of 10$\mu$m thickness, which were 
tagged with the upgraded Glasgow magnetic spectrometer \cite{Anthony_91,Hall_96,McGeorge_08} 
for photon energies from 0.45 GeV to 1.4 GeV. The typical bin width for the photon 
beam energy (4 MeV) is defined by the geometrical size of the plastic scintillators in the 
focal plane detector of the tagger. The intrinsic resolution of the magnetic spectrometer 
is better by more than an order of magnitude. The size of the tagged photon beam
spot on the target was restricted by a 3 mm diameter collimator placed downstream 
from the radiator foil. 
The target was a mylar cylinder of 3.0~cm diameter and 5.08~cm
length filled with liquid $^3$He at a temperature of 2.6~K. The corresponding
target density was 0.073 nuclei/barn.

The reaction products were detected with an electromagnetic calorimeter combining
the Crystal Ball detector (CB) \cite{Starostin_01} made of 672 NaI crystals with 
384 BaF$_2$ crystals from the TAPS detector \cite{Novotny_91,Gabler_94}, configured 
as a forward wall. The Crystal Ball was equipped with an additional 
{\bf P}article {\bf I}dentification {\bf D}etector (PID) \cite{Watts_04} for the 
identification of charged particles and all modules of the TAPS detector had individual
plastic scintillators in front for the same purpose. 
The PID, in combination with 
the Crystal Ball, and the TAPS - TAPS-Veto system could be used for a $E- \Delta E$
analysis for the separation of different charged particles.
The Crystal Ball covered the full azimuthal range for polar 
angles from 20$^{\circ}$ to 160$^{\circ}$, corresponding to 93\% of the full solid angle.
The TAPS detector, mounted 1.457 m downstream from the target, covered 
polar angles from $\approx$5$^{\circ}$ to 21$^{\circ}$. This setup is similar to 
the one described in detail in \cite{Schumann_10}. The only difference is that, in the 
earlier setup, the TAPS forward wall consisted of 510 modules and was placed 1.75~m 
from the target. 

For the present experiment the main trigger condition was a multiplicity of two
separated hits in the combined CB/TAPS system and an integrated energy deposition 
of at least 300 MeV in the Crystal Ball. Details for the basis of the data analysis 
including calibration procedures, identification of photons and recoil nucleons 
and the absolute normalization of cross sections are discussed
in \cite{Schumann_10}. Here, we will only outline the specific
steps for the identification of coherent $\eta$-production. Results for quasi-free
production of $\eta$-mesons off nucleons bound in $^3$He in view of the structure
recently observed for $\eta$-photoproduction off the neutron
\cite{Jaegle_08,Jaegle_11} will be reported elsewhere. 

The analysis was based on the $\eta\rightarrow \gamma\gamma$ and  
$\eta\rightarrow 3\pi^0\rightarrow 6\gamma$ decay channels. The measurement of both
channels allowed an additional control of systematic uncertainties.
Events were analyzed with either two or six photon candidates and no further
hit in the detector system. The latter condition reduces the incoherent background
since events with recoil nucleons were suppressed. For the two-photon decay,
$\eta$-mesons were identified with a standard invariant-mass analysis.
Residual background under the $\eta$-peak was subtracted by fitting the spectra
with simulated line shapes and a background polynomial. This procedure was applied 
individually for each combination of incident photon energy and meson c.m.-polar
angle. For the six photon decay, the photons were first combined via a $\chi^2$-test
to three pairs, which were the best solution for three $\pi^0$ invariant masses.
A cut between 110 MeV - 150 MeV was made on these invariant masses. Subsequently,
the six-photon invariant mass was constructed. The corresponding spectra were 
basically background free under the $\eta$-peak (direct triple-$\pi^0$ production
has a very small cross section in the energy range of interest).

The most important step in the analysis is the separation of the coherent reaction 
from breakup where nucleons are removed from the nucleus. 
Since the $^3$He recoil nuclei cannot be detected (they are mostly stopped in the target)
only the overdetermined reaction kinematics can be used. 
The kinetic energy of the
$\eta$-mesons in the $\gamma$-$^3$He c.m.-system is fixed not only by the incident 
photon energy but also from the measured laboratory energy and polar angle of the meson. 
\begin{figure}[htb]
\resizebox{0.49\textwidth}{!}{%
  \includegraphics{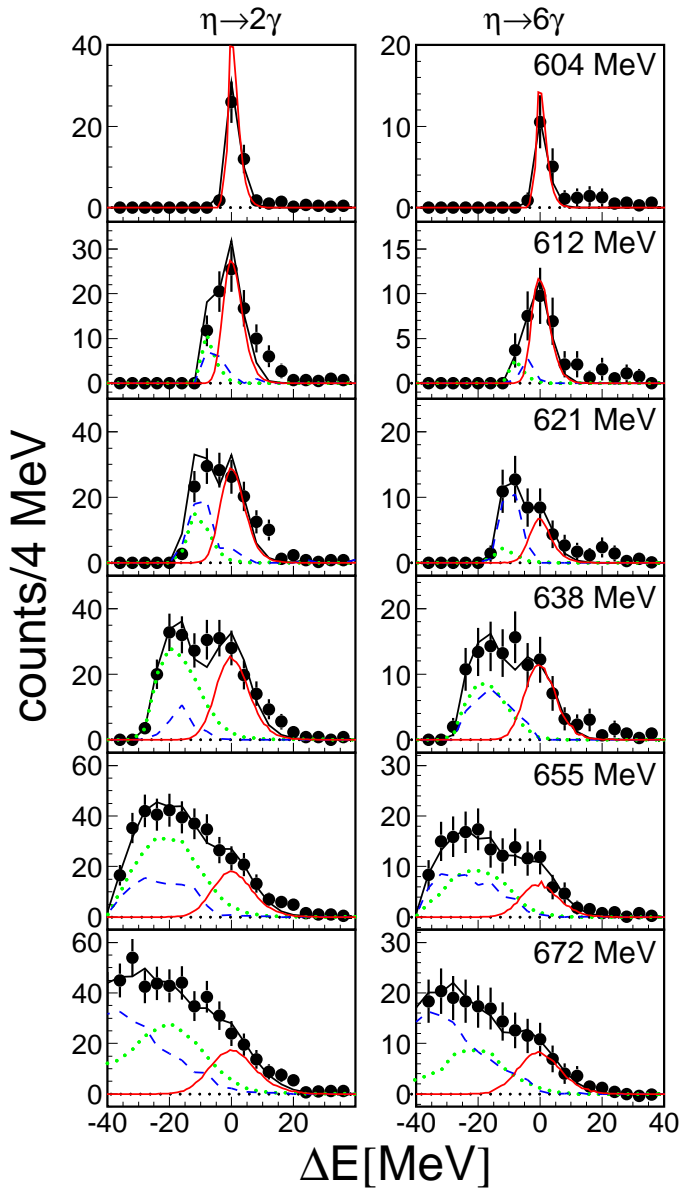}
}
\caption{Missing energy spectra for the $\gamma ^3\mbox{He}\rightarrow \eta^3\mbox{He}$
reaction for the two-photon and 3$\pi^0$ decay modes for different ranges of incident
photon energy. Black dots: experiment. Curves: Monte Carlo simulations. Solid (red): coherent
contribution, dashed (blue): recoil taken by quasi-free nucleon, dotted (green): recoil
taken by di-nucleon, solid (black) sum of all.
}
\label{fig:misse}       
\end{figure}
The difference, the missing energy $\Delta E$, is shown for both data samples in 
Fig.~\ref{fig:misse} and compared to the simulated line shapes of the coherent 
and breakup reactions. The simulations were done with the Geant4 
program package \cite{GEANT4}, taking into account all details of the target and 
detector setup. The simulation of the coherent part is straightforward due to the 
two-body final-state kinematics. In case of the breakup part, the momentum 
distribution of the bound nucleons has to be considered, and this was
taken from the work of McCarthy, Sick, and Whitney \cite{McCarthy_77}. 
Several final states, such as $pd$, $npp$ with different participant - spectator
combinations, contribute. Good agreement between data and simulations was
achieved, at all the incident photon energies investigated, with three different 
reaction components: the coherent process, the breakup process with a quasi-free 
participant nucleon, and the breakup process where the recoil is taken by a `participant'
deuteron. 

The relative contributions of these processes have been determined by
fitting the data with a superposition of their line shapes (cf Fig.~\ref{fig:misse}).
Since in principle the detection efficiency for the different components
can be different for the $\eta\rightarrow 2\gamma$ and $\eta\rightarrow 6\gamma$
decays, no constraints relating the two channels were imposed on the fits.
As a consequence, for some energy bins the fitted relative contributions of the 
breakup processes are different for the two decay channels, but the fit of the coherent 
contribution is quite stable. This is so because the breakup background makes almost no 
contribution at positive missing energies, so that the coherent contribution is only 
weakly dependent on the exact shape of the background. The lowest range of photon 
energies (600 MeV $< E_{\gamma} <$ 608 MeV) lies between the kinematic thresholds of 
coherent and breakup reactions. Consequently, only the coherent reaction can 
contribute. In fact, a clean signal at zero missing energy is seen, which agrees 
with the simulated coherent signal shape. At even lower photon energies, the spectra 
show no signal above statistical noise. 

The count rates are roughly a factor of
2.75 larger for the $\eta\rightarrow \gamma\gamma$ data than for the 
$\eta\rightarrow 3\pi^0\rightarrow 6\gamma$ data (note the different scales for the
left and right parts of Fig.~\ref{fig:misse}), which is due to the respective
detection efficiencies ($\approx$80\% for $2\gamma$, $\approx$35\% for $6\gamma$)
and decay branching ratios ($b_{2\gamma}$=(39.31$\pm$0.20)\%, 
$b_{6\gamma}$=(32.57$\pm$0.23)\% \cite{PDG}). 
Absolutely normalized total and differential cross sections for both decay 
channels have been extracted from the yields, the target density, the decay branching 
ratios \cite{PDG}, the simulated detection efficiency, the electron beam flux measured
in the focal-plane detector of the tagging spectrometer, and the tagging efficiency
$\epsilon_{\rm tag}$, i.e. the number of correlated photons that pass 
through the collimator. 
The latter was measured in special low-intensity runs with a lead-glass detector 
in the photon beam (see \cite{Schumann_10} for details) and ranged from
60\% to 75\%. For the systematic uncertainties we estimate 10\% for the overall
normalization including target density (measurement of target temperature and 
possible deformation of the cold target cell, $\approx$7-8 \%), photon flux (measurement 
of tagging efficiency and electron flux, $<$5\%) and decay branching ratios 
(almost negligible $<$1 \%). For the reaction dependent simulations of the
detection efficiency, 5\% uncertainties are estimated. Finally, the energy dependent
uncertainty for the reaction identification (in particular the separation in the
missing energy spectra) ranges from 2\% at threshold to 20\% at the highest incident
photon energies.

For the analysis of the excitation function of $\pi^0$-p back-to-back pairs in
the $\gamma$-$^3$He c.m.-system, Monte Carlo simulations were done for the 
signal and the background coming from quasi-free production of $\pi^0$ mesons. 
For the signal, the decay of bound S$_{11}$ resonances into $N\pi$ was modelled with 
a momentum distribution corresponding to the nuclear Fermi-motion.
These simulations were used to select the kinematical conditions
best suited for the signal. In the spectra of the pion-proton opening angle,
signal events appear roughly between 150$^{\circ}$ - 180$^{\circ}$, while the background 
is distributed between 100$^{\circ}$ and 180$^{\circ}$.

\section{Results}
\subsection{Coherent photoproduction of $\eta$-mesons off $^3$He}

\begin{figure}[thb]
\resizebox{0.47\textwidth}{!}{%
  \includegraphics{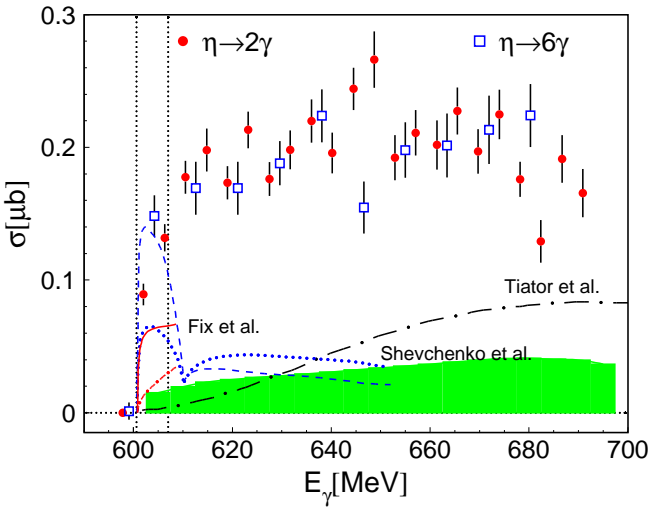}
}
\caption{Total cross section for the $\gamma ^3$He$\rightarrow$$\eta ^3$He
reaction from $\eta\rightarrow 2\gamma$ and $\eta\rightarrow 6\gamma$ decays.
The shaded band at the bottom indicates the systematic uncertainty. The two vertical
lines indicate coherent and breakup threshold.
Theory curves: (blue) dotted and dashed
from Shevchenko et al. \cite{Shevchenko_03} for two different versions of elastic
$\eta N$ scattering, (red) solid (dash-dotted): Fix and Arenh\"ovel \cite{Fix_03}
full model (plane wave), (black) long dash-dotted: Tiator et al. \cite{Tiator_94}. 
}
\label{fig:total_raw}       
\end{figure}

The total cross section data for coherent $\eta$-production from the two $\eta$
decay branches is compared to model predictions \cite{Tiator_94,Fix_03,Shevchenko_03} 
in Fig.~\ref{fig:total_raw}. The agreement between the two data sets is quite good, but 
none of the existing models reproduce the data, even when their systematic 
uncertainties (shown as shaded band in the figure) are considered. The most prominent
feature of the data is the extremely sharp rise at threshold. Here one should note that
the binning of the data (4 MeV for the 2$\gamma$-channel, 8 MeV for the 6$\gamma$-channel)
is very large compared to the energy resolution of the magnetic tagging spectrometer
(roughly 200 keV).

The work by Tiator, Bennhold and Kamalov \cite{Tiator_94} is based
on a coupled-channel analysis of pion- and photon-induced pion- and eta-production off
the nucleon, which parameterizes the resonance contributions with an isobar model and 
the background terms with effective Lagrangians. For the coherent production off
$^3$He, realistic nuclear wave functions were used, but the model was based on
the plane-wave impulse approximation (PWIA). The result strongly underestimates the
magnitude of the measured cross section and does not reproduce the energy dependence.    

Fix and Arenh\"ovel \cite{Fix_03} have modelled coherent $\eta$-photoproduction 
off $^3$He and $^3$H in PWIA, in a distorted-wave impulse approximation (DWIA), using
an optical potential, and in a full four-body scattering model. They find strong
FSI effects, which amplify the threshold cross section for the full four-body model 
with respect to PWIA and DWIA (which give similar results). Nevertheless their
cross section underestimates the data by roughly a factor of two. 

Shevchenko and
collaborators \cite{Shevchenko_03} have also calculated coherent $\eta$-production off
the three-nucleon system in a microscopic few-body description. They find a strong
dependence of the result on the elastic $\eta N$ rescattering, which is not sufficiently
constrained by experiment. Two examples for different FSI modelling are shown in 
Fig.~\ref{fig:total_raw}. They exhibit strong threshold effects, but do not
reproduce the measurements above the breakup threshold.

\begin{figure}[thb]
\resizebox{0.47\textwidth}{!}{%
  \includegraphics{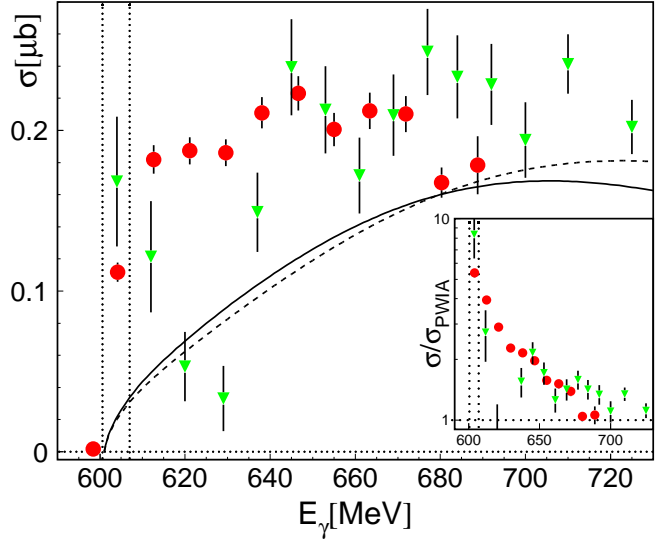}
}
\caption{Total cross section for $\gamma ^3$He$\rightarrow$$\eta ^3$He
(averaged over $2\gamma$ and $3\pi^0$ decays) (red dots) compared to previous data 
\cite{Pfeiffer_04} (green triangles). Solid (dashed) curves: 
PWIA with realistic (isotropic) angular distribution for 
$\gamma n\rightarrow n\eta$ (see text). The present data are binned in the same way 
as the angular distributions in Fig. \ref{fig:ang} (bin width $\approx$8~MeV).
Insert: ratio of measured and PWIA cross sections. 
}
\label{fig:total}       
\end{figure}

The average of the two data sets is compared to the previous result from 
Pfeiffer et al. \cite{Pfeiffer_04} in Fig.~\ref{fig:total}. The two data sets are in 
reasonable agreement (within uncertainties) except for two points 
around 620 MeV. Here, one should note that, in this range, the systematic uncertainty in 
the previous data was large because the coherent signal had to be extracted by fitting a small
coherent contribution in the missing energy spectra, which were dominated by the quasi-free 
reaction. The quasi-free contribution is less important at lower incident photon energies, 
and is better separated from the coherent component at higher incident photon energies.
The present experiment profited in this range from the almost $4\pi$ coverage of the detector, 
which allowed some suppression of the quasi-free background by the detection of associated 
recoil nucleons.

Also shown in this figure 
are the results from a simple PWIA model. It uses the effective photon energy 
$E_{\gamma}^{\mbox{eff}}(E_{\gamma},x)$ given by 
\begin{equation}
E_{\gamma}^{\mbox{eff}} = \frac{s_{\mbox{eff}}-m_N^2}{2m_N},\;\;\;\;\;\;\;
s_{\mbox{eff}} = (P_{\gamma} + P_{N})^2,
\end{equation}
where $E_{\gamma}$ is the laboratory energy of the incident photon and 
$\Theta_{\eta}^{\star}$ is the polar angle of the $\eta$ meson in the photon-nucleus 
c.m. system with $x=\mbox{cos}(\Theta_{\eta}^{\star})$. The effective photon energy
corresponds to the $s_{\mbox{eff}}$ of the incident photon (four-momentum $P_{\gamma}$)
and a nucleon (four-momentum $P_{N}$) with three-momentum ${\bf{p}}_N$ from the nucleon 
motion inside the nucleus. The nucleon momentum is related in the factorization 
approximation to the momentum $\bf{q}$ transferred to the nucleus by \cite{Chumbalov_87}:    
\begin{equation}
{\bf{p}}_N = -\frac{A-1}{2A}{\bf{q}} = -\frac{1}{3}{\bf{q}}\; ,
\end{equation}
where all momenta are in the laboratory system.

The coherent cross section is then composed of three factors:
\begin{equation}
\frac{d\sigma_{\mbox{PWIA}}}{d\Omega}(E_{\gamma},x) = 
\left(\frac{q_{\eta}^{(A)}}{k_{\gamma}^{(A)}}
\cdot\frac{k_{\gamma}^{(N)}}{q_{\eta}^{(N)}}\right) 
\cdot\frac{d\sigma_{\mbox{elem}}}{d\Omega}
\cdot\frac{F^2_A(q^2)}{F^2_p(q^2)}
\label{eq:main}
\end{equation}
the elementary cross section $d\sigma_{\mbox{elem}}/d\Omega$, a kinematical factor, and
the nuclear form factor for point-like nucleons. The kinematical factor accounts for 
the change of phase-space between the different c.m. systems and is derived from the photon 
and $\eta$ three-momenta in the photon-nucleon ($k_{\gamma}^{(N)}$, $q_{\eta}^{(N)}$), 
and photon-nucleus ($k_{\gamma}^{(A)}$, $q_{\eta}^{(A)}$) c.m. systems. 
The nuclear form-factor $F^2_A(q^2)$ was taken from \cite{McCarthy_77}.
It includes the spatial distribution of the charge (respectively magnetic moment) 
of the nucleon, while here the distribution of point-like nucleons is relevant.
We have therefore used the ratio of the nuclear form factor $F_{A}(q^2)$ and the nucleon
dipole form factor $F_{p}(q^2)$ to approximate the distribution of point-like nucleons in
the $^3$He nucleus.  

\begin{figure}[htb]
\resizebox{0.47\textwidth}{!}{%
  \includegraphics{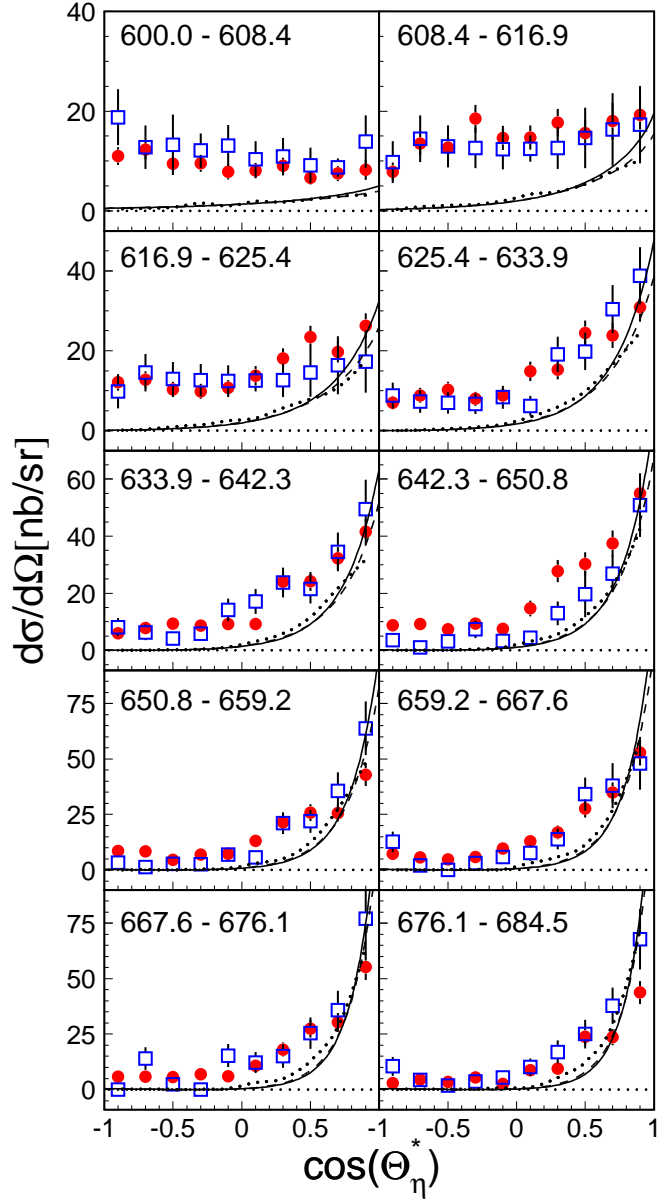}
}
\caption{Angular distributions of $\gamma ^3$He$\rightarrow$$\eta ^3$He
in the photon-nucleon c.m. system for different energy bins (range of incident photon
energy in MeV indicated by the labels). 
(Red) dots: $\eta\rightarrow 2\gamma$ decay, 
(blue) open squares: $\eta\rightarrow 6\gamma$ decay, dashed curves:
results of PWIA model with isotropic angular distributions for 
$\gamma n\rightarrow n\eta$, solid curves with realistic angular distributions,
dotted curves: folded with experimental resolution.
(see text).
}
\label{fig:ang}       
\end{figure}

Since $\eta$-photoproduction in this energy range is almost only due to the $S_{11}$
excitation, the dominant electromagnetic multipole is the $E_{0^+}$ spin-flip.
Neglecting small contributions from other multipoles and higher components in the 
$^3$He wave function, only the unpaired neutron contributes to coherent production
because a spin-flip of a proton is Pauli-blocked. Therefore, we use for the elementary
cross section the experimental results for the $\gamma n\rightarrow n\eta$ reaction
from \cite{Jaegle_08,Jaegle_11}. Since the angular dependence up to the S$_{11}$
peak is almost isotropic, one can approximate the c.m. differential cross section by:
\begin{equation}
\frac{d\sigma_{\mbox{elem}}}{d\Omega}(E_{\gamma}^{\mbox{eff}},x) = 
\sigma_n(E_{\gamma}^{\mbox{eff}})/4\pi\;,
\label{eq:isotropic}
\end{equation}
where the total neutron cross section $\sigma_{n}$ is parameterized as a Breit-Wigner
curve with an energy dependent width using the numerical values from \cite{Jaegle_11}
($W$=1546 MeV, $\Gamma$=176 MeV, $A^n_{1/2}$=90$\times$10$^{-3}$ GeV$^{-1/2}$,
$b_{\eta}$=0.5, $b_{\pi}$=0.4, $b_{\pi\pi}$=0.1).
Variation of the parameters within their uncertainties changes the predictions for 
the coherent cross section typically on the 10\% level. 

The results from the PWIA model are compared to the data in Figs.~\ref{fig:total}
and \ref{fig:ang}. They are shown for the approximation of Eq.~\ref{eq:isotropic} 
assuming isotropic angular distributions for the elementary
$\gamma n\rightarrow n\eta$ reaction, and also for realistic angular distributions
taken from \cite{Jaegle_11}. The difference is, as expected, negligible.
Despite the simplicity of the model, the predicted cross section agrees with experiment
within statistical and systematic uncertainties for the highest incident photon energies. 
In the threshold  region it is in excellent agreement with the PWIA calculation by Fix and 
Arenh\"ovel. However, in this region it underestimates the data by nearly one order of 
magnitude, which is evidence for strong FSI effects at threshold. 

The behavior of the angular distributions of the PWIA is dominated by the
nuclear form factor, which is responsible for the strong forward peaking.  
At high incident photon energies the data show the same tendency, though the rise to 
forward angles is not as steep as in the model. This indicates significant FSI effects
even for those energies, since the more trivial approximations in the PWIA model
cannot explain this discrepancy. The use of realistic angular distributions for
the elementary cross section instead of Eq.~\ref{eq:isotropic} 
(dashed curves in Fig. \ref{fig:ang}) has basically no impact. The use of more
realistic nuclear wave functions instead of the simplified scalar form factor in 
Eq.~\ref{eq:main} leads only to comparatively small changes \cite{Fix_03}. 
Finally, the experimental angular resolution is only responsible for minor effects.
This is demonstrated by the dotted curves in Fig. \ref{fig:ang}, which have been folded 
with the detector response.   

Towards the threshold, the measured angular distributions become almost isotropic.
Between coherent and breakup threshold, they develop a rise to backward angles, 
while in PWIA the form factor still causes a forward peaking. Together with the
disagreement in the absolute scale of the cross section in the threshold region, 
this comparison clearly demonstrates the large influence of FSI effects.

\subsection{Excitation function of $\pi^0-p$ back-to-back pairs}

\begin{figure}[htb]
\resizebox{0.49\textwidth}{!}{%
  \includegraphics{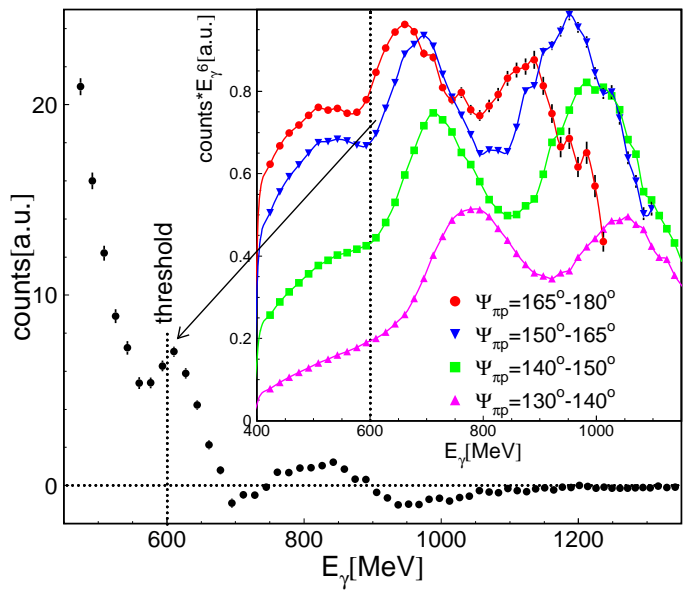}
}
\caption{Main plot: difference of excitations functions of $\pi^0-p$ back-to-back 
pairs with opening angles between 165$^{\circ}$ - 180$^{\circ}$ and 
150$^{\circ}$ - 165$^{\circ}$. Insert excitations functions for different ranges 
of the opening angle $\Psi_{\pi p}$ after removal of the overall energy dependence 
$\propto E_{\gamma}^{-6}$. Vertical dotted lines: coherent $\eta$-production threshold.
}
\label{fig:pi0}       
\end{figure}

The excitation function of $\pi^0$-proton pairs for different ranges of their opening
angle in the $\gamma$-$^3$He c.m. system was analyzed in a manner similar to that of 
Pfeiffer et al. \cite{Pfeiffer_04}. The idea was, that when $\eta$-mesons are 
produced into an $\eta$-$^3$He resonant state, overlapping with the coherent production 
threshold, those produced at photon energies below the threshold
are off-shell and cannot be emitted without violating energy conservation. They can, 
however, be captured by a nucleon which is then excited to the S$_{11}$-resonance.
Since this resonance has a $\approx$50\% decay branching ratio into $N\pi$, the expected 
signal would be pion-nucleon pairs emitted back-to-back in the $\gamma$-$^3$He c.m. system. 
The results are shown in Fig.~\ref{fig:pi0}.
The figure shows the difference of the excitation functions for opening angles from 
165$^{\circ}$ - 180$^{\circ}$ (back-to-back range taking into account effects of Fermi 
motion) and the data from 150$^{\circ}$ - 165$^{\circ}$. These ranges were optimized
with Monte Carlo simulations of signal and background and are slightly
different from those used by Pfeiffer et al. (170$^{\circ}$-180$^{\circ}$ and 
160$^{\circ}$-170$^{\circ}$), but this is a minor effect.
The result shows the peak at the coherent $\eta$ threshold 
that had been observed previously, but now with much higher statistical 
significance. However, the data also show some structure at higher incident photon 
energies which was not previously seen because the earlier experiment covered only 
incident photon energies up to 800 MeV. The much higher statistical quality of the 
new data allowed a detailed analysis of these structures. This is summarized in the 
insert of Fig. \ref{fig:pi0}.

It shows the excitation functions for different ranges of opening angle, when the 
overall energy dependence of the data ($\propto E_{\gamma}^{-6}$) is removed. 
The broad, peak-like structures are due to the second and third resonance regions 
of the nucleon. The position of these signals shifts to higher incident photon 
energies for smaller opening angles. This is a purely kinematical effect;
small opening angles correspond to decays of nucleon resonances moving forward
in the photon-nucleus c.m. system. Therefore, for a given mass $M_R=W$ of the resonance,
small opening angles correspond to large incident photon energies (and nucleon 
Fermi momenta parallel to the photon momentum). Unfortunately, subtraction of the
normalized excitation functions for the 165$^{\circ}$ - 180$^{\circ}$ and 
150$^{\circ}$ - 165$^{\circ}$ opening-angle ranges produces a narrow peak exactly
at the $\eta$-production threshold. This peak does, however, not originate from a
special structure in the back-to-back data, but it is due to the subtraction of the 
shifted low energy tails of the second resonance region in single pion production.

Here, we have only presented the results from an analysis similar to the previous 
work by Pfeiffer et al. \cite{Pfeiffer_04}. The much 
better statistical quality of the present data allows also more refined analyses
with additional cuts to enhance the signal-to-background ratio. For this purpose we
have done detailed Monte Carlco simulations of the kinematical correlations for the 
signal and the background from quasi-free pion production and analyzed the 
excitation functions with optimized cuts. However, also with these analyses no
structures that could be uniquely related to the decay of an $\eta$-mesic state
into $N\pi$ could be identified. The $\pi^0 - p$ background before very specific 
cuts is at least three orders of magnitude larger than any expected signal from an
$\eta$-mesic state. In consequence, it seems to be impossible to identify in this 
reaction channel a small signal from the decay of bound S$_{11}$ resonances on top of 
these complicated background structures, even when the signal exists.

\section{Summary and conclusions}

Coherent photoproduction of $\eta$-mesons off $^3$He has been measured with much
improved statistical quality compared to the pilot experiment of Pfeiffer et al.
\cite{Pfeiffer_04}. The total cross section rises sharply between the coherent and
breakup thresholds. 
Compared to a PWIA, which is in fair agreement with the data in the
S$_{11}$ peak, the threshold values are enhanced by nearly one order of magnitude.
This is very different e.g. from the behavior of coherent $\eta$ photoproduction off
the deuteron, which is in reasonable agreement with PWIA \cite{Weiss_01}.
The angular distributions at threshold are almost isotropic, and are unlike the forward 
peaked distributions expected to result from the form factor behavior.
This result is similar to that previously observed for the hadron induced
reactions $pd\rightarrow\eta  ^3\mbox{He}$ \cite{Mayer_96} and
$dp\rightarrow\eta  ^3\mbox{He}$ \cite{Smyrski_07,Mersmann_07,Rausmann_09}.
This independence from the initial state is strong evidence for dominant $\eta$-nucleus
interaction effects, related to a resonant state in the vicinity of the $\eta$ production 
threshold.

The excitation function of $\pi^0-p$ back-to-back pairs was investigated
as an independent signal for the formation of a quasi-bound or virtual
$\eta$-nucleus state, as originally suggested in the work by Sokol et al. 
\cite{Sokol_99,Sokol_08}. A peak-like structure at the $\eta$-production threshold had
been observed in a previous measurement of photoproduction from $^3$He by Pfeiffer et
al. \cite{Pfeiffer_04}. This signal was reproduced with much improved statistical
significance, but it could be attributed to an artifact arising from the complicated
background structure of quasi-free pion production.

\vspace*{0.5cm}
{\bf Acknowledgments}

We wish to acknowledge the outstanding support of the accelerator group 
and operators of MAMI. We gratefully thank C. Wilkin for many stimulating discussions,
instructive suggestions, and for careful and critical reading of the manuscript.
We thank A. Fix and L. Tiator for very helpful discussions about the PWIA
calculations.
This work was supported by Schweizerischer Nationalfonds, Deutsche
Forschungsgemeinschaft (SFB 443, SFB/TR 16), DFG-RFBR (Grant No. 05-02-04014),
UK Science and Technology Facilities Council, STFC, European Community-Research 
Infrastructure Activity (FP6), the US DOE, US NSF and NSERC (Canada).


\begin{thebibliography}{99}
\bibitem{Geissel_02}            H. Geissel, et al.,                              Phys. Rev. Lett.                    88    (2002)    122301.
\bibitem{Krusche_95}            B. Krusche et al.,                                Phys. Rev. Lett.                    74   (1995)      3736.
\bibitem{Krusche_97}            B. Krusche et al.,                                Phys. Lett.                      B 397   (1997)       171.
\bibitem{PDG}                   K. Nakamura et al.,                               J. Phys.                         G  37   (2010)    075021.
\bibitem{Roebig_96}             M. R\"obig-Landau et al.,                         Phys. Lett.                      B 373   (1996)        45.
\bibitem{Mertens_08}            T. Mertens et al.,                                Eur. Phys. J.                    A  38   (2008)       195.
\bibitem{Bhalerao_85}           R.S.Bhalerao and L.C.Liu,                         Phys. Rev. Lett.                    54   (1985)       865.
\bibitem{Liu_86}                L.C.Liu and Q.Haider,                             Phys. Rev.                       C  34   (1986)      1845.
\bibitem{Chrien_88}             R.E.Chrien et al.,                                Phys. Rev. Lett.                    60   (1988)      2595.
\bibitem{Johnson_93}            J.D.Johnson et al.,                               Phys. Rev.                       C  47   (1993)      2571.
\bibitem{Sokol_99}              G.A. Sokol et al.,                                Fizika                           B   8   (1999)        85.
\bibitem{Sokol_08}              G.A. Sokol and L.N. Pavlyuchenko,                 Phys. of. Atomic Nuclei             71   (2008)       509.
\bibitem{Krusche_95a}           B. Krusche at al.,                                Phys. Lett.                      B 358   (1995)	 40.
\bibitem{Hoffmann_97}           P. Hoffmann-Rothe et al.,                         Phys. Rev. Lett.                    78   (1997)      4697.
\bibitem{Weiss_01}              J. Weiss et al.,                                  Eur. Phys. J.                    A  11   (2001)	371.
\bibitem{Weiss_03}              J. Weiss et al.,                                  Eur. Phys. J.                    A  16   (2003)	275.
\bibitem{Hejny_99}              V. Hejny et al.,                                  Eur. Phys. J.                    A   6   (1999)	 83.
\bibitem{Hejny_02}              V. Hejny et al.,                                  Eur. Phys. J.                    A  13   (2002)	493.
\bibitem{Arndt_05}              R.A. Arndt et al.,                                Phys. Rev.                       C  72   (2005)    045202. 
\bibitem{Ueda_91}               T.Ueda,                                           Phys. Rev. Lett.                    66   (1991)       297.
\bibitem{Ueda_92}               T.Ueda,                                           Phys. Lett.                      B 291   (1992)       228.
\bibitem{Wilkin_93}             C.Wilkin,                                         Phys. Rev.                       C  47   (1993)      R938.
\bibitem{Rakityanski_95}        S.A.Rakityanski et al.,                           Phys. Lett.                      B 359   (1995)        33. 
\bibitem{Rakityanski_96}        S.A. Rakityanski et al.,                          Phys. Rev.                       C  53   (1996)     R2043.
\bibitem{Green_96}              A.M. Green and S. Wycech,                         Phys. Rev.                       C  54   (1996)      1970.
\bibitem{Scoccola_98}           N.N. Scoccola, D.O. Riska,                        Phys. Lett.                      B 444   (1998)        21. 
\bibitem{Shevchenko_00}         N.V. Shevchenko et al.,                           Eur. Phys. J.                    A   9   (2000)       143.
\bibitem{Grishina_00}           V. Yu. Grishina et al.,                           Phys. Lett.                      B 475   (2000)         9. 
\bibitem{Garcilazo_01}          H. Garcilazo and M.T. Pe\~{n}a,                   Phys. Rev.                       C  63   (2001) 021001(R).
\bibitem{Peng_87}               J.C. Peng et al.,                                 Phys. Rev. Lett.                    58   (1987)      2027.
\bibitem{Peng_89}               J.C. Peng et al.,                                 Phys. Rev. Lett.                    63   (1989)      2353.
\bibitem{Calen_96}              H. Cal\'en et al.,                                Phys. Lett.                      B 366   (1996)       366. 
\bibitem{Smyrski_00}            J. Smyrski et al.,                                Phys. Lett.                      B 474   (2000)       182. 
\bibitem{Moskal_04}             P. Moskal  et al.,                                Phys. Rev.                       C  69   (2004)    025203.
\bibitem{Plouin_90}             F. Plouin, P.Fleury, and C. Wilkin,               Phys. Rev. Lett.                    65   (1990)       690. 
\bibitem{Calen_98}              H. Cal\'en et al.,                                Phys. Rev. Lett.                    80   (1998)      2069. 
\bibitem{Mayer_96}              B. Mayer et al.,                                  Phys. Rev.                       C  53   (1996)      2068. 
\bibitem{Smyrski_07}            J. Smyrski et al.,                                Phys. Lett.                      B 649   (2007)       258.
\bibitem{Mersmann_07}           T. Mersmann et al.,                               Phys. Rev. Lett.                    98   (2007)    242301.
\bibitem{Rausmann_09}           T. Rausmann et al.,                               Phys. Rev.                       C  80   (2009)    017001.
\bibitem{Wronska_05}            A. Wronska et al.,                                Eur. Phys. J                     A  26   (2005)       421.
\bibitem{Willis_97}             N. Willis et al.,                                 Phys. Lett.                      B 406   (1997)        14.
\bibitem{Budzanowski_09}        A. Budzanowski et al.,                            Nucl. Phys.                      A 821   (2009)       193.
\bibitem{Hibou_00}              F. Hibou et al.,                                  Eur. Phys. J.                    A   7   (2000)       537. 
\bibitem{Wilkin_07}             C. Wilkin et al.,                                 Phys. Lett.                      B 654   (2007)        92.
\bibitem{Krusche_03}            B. Krusche and S. Schadmand,                      Prog. Part. Nucl. Phys.             51   (2003)       399. 
\bibitem{Pfeiffer_04}           M. Pfeiffer et al.,                               Phys. Rev. Lett.                    92   (2004)    252001. 
\bibitem{Sibirtsev_04}          A. Sibirtsev et al.,                              Phys. Rev.                       C  70   (2004)    047001.
\bibitem{Hanhart_05}            C. Hanhart,                                       Phys. Rev. Lett.                    94   (2005)    049101.
\bibitem{Herminghaus_83}        H. Herminghaus et al.,                            IEEE Trans. on Nucl. Science.       30   (1983)      3274.
\bibitem{Kaiser_08}             K.-H. Kaiser et al.,                              Nucl. Instr. Meth.               A 593   (2008)	159.
\bibitem{Anthony_91}            I. Anthony et al.,                                Nucl. Instr. Meth.               A 301   (1991)	230.
\bibitem{Hall_96}               S.J. Hall, G.J. Miller, R. Beck, P.Jennewein,     Nucl. Inst.Meth.                 A 368   (1996)       698.
\bibitem{McGeorge_08}           J.C. McGeorge et al.,                             Eur. Phys. J.                    A  37   (2008)       129.
\bibitem{Starostin_01}          A. Starostin et al.,                              Phys. Rev.                       C  64   (2001)    055205.
\bibitem{Novotny_91}            R. Novotny,                                       IEEE Trans. on Nucl. Science        38   (1991)	379. 
\bibitem{Gabler_94}             A.R. Gabler et al.,                               Nucl. Instr. Meth.               A 346   (1994)       168.
\bibitem{Watts_04}              D. Watts, in {\em Calorimetry in Particle Physics, Proceedings of the 11th International Conference, Perugia,
                                Italy 2004}, edited by C. Cecchi, P. Cenci, P. Lubrano, and M. Pepe (World Scientific, Singapore, 2005, p. 560 
\bibitem{Schumann_10}           S. Schumann et al.,                               Eur. Phys. J.                    A  43   (2010)       269.
\bibitem{Jaegle_08}             I. Jaegle et al.,                                 Phys. Rev. Lett.                   100   (2008)    252002.
\bibitem{Jaegle_11}             I. Jaegle et al.,                                 Eur. Phys. J.                    A  47   (2011)        89. 
\bibitem{GEANT4}                S. Agostinelli et al.,                            Nucl. Instr. Meth.               A 506   (2003)       250. 
\bibitem{McCarthy_77}           J. S. McCarthy, I. Sick, R. R. Whitney,           Phys. Rev.                       C  15   (1977)      1396.
\bibitem{Tiator_94}             L. Tiator, C. Bennhold, and S.S. Kamalov          Nucl. Phys.                      A 580   (1994)       455.;
and S.S. Kamalov, priv. com. (2004).
\bibitem{Fix_03}                A. Fix and H. Arenh\"ovel,                        Phys. Rev.                       C  68   (2003)    044002.;
and A. Fix, priv. com. (2011).
\bibitem{Shevchenko_03}         N.N. Shevchenko et al.,                           Nucl. Phys.                      A 714   (2002)       277.
\bibitem{Chumbalov_87}          A.A. Chumbalov, R.A. Eramzhyan, and S.S. Kamalov, Z. Phys.                         A 328   (1987)       195.;
and L. Tiator, priv. com. (2011).
\end{thebibliography}
\end{document}